# LINUX-BASED TOOLKIT ON THE VEPP-4 CONTROL SYSTEM

V. Blinov, A. Bogomyagkov, M. Fedotov, S. Karnaev, O. Meshkov, N. Muchnoi, S. Nikitin,
I. Nikolaev, A. Selivanov, S. Smirnov, V. Tsukanov,
Budker Institute of Nuclear Physics, Novosibirsk

## Abstract

The control system of the VEPP-4 facility was designed more than fifteen years ago and based on the home-developed CAMAC-embedded minicomputers Odrenok [1]. Five years ago, all computers were connected via Ethernet network. This step allowed us to force an integration of PCs into the VEPP-4 control system. This paper reviews new tools running on the PCs under Linux.

The beam diagnostics tool described in this paper provides the data about position and dimensions of $e^-/e^+$ beams from CCD matrix controller via 100-Mbit Ethernet.

The next tool provides measuring of beam energy at the VEPP-4M collider using a well-known method of resonance depolarization by the observation of the polarization degree on the effect of internal scattering of particles.

Data visualization tools are based on CERN ROOT framework.

Hardware and software aspects of the systems are presented in this paper.

## 1 INTRODUCTION

The first steps to control accelerators in BINP with computers began at the end of the 1960's when ODRA-1304 machine (ICL-1900 series) was used to control the acceleration process on the VEPP-3 storage ring. Later in the 1970's, several ODRA-1305 and ODRA-1325 machines were implemented at the VEPP-4 complex, which includes injection complex, VEPP-3 (2 GeV electron/positron storage ring), and VEPP-4M (1 – 6 GeV electron-positron collider). During this time a great number of electronic modules, applications, and a custom real-time operation system were developed in BINP. In the 1980's the VEPP-4 control system was upgraded. The main goal was to replace obsolete equipment and electronics by CAMAC. In the beginning of the 1980's Odernok intelligent CAMAC controller with ODRA instruction set was developed in BINP. The use of Odrenok allowed to maintain customary programming for more than 20 years. All applications were programmed on the ODRA and Odrenok in the custom version of Fortran language – TRAN 1900. In 1998 a custom Ethernet network was implemented for Odrenok [2]. This effort led to the use of PCs as operator consoles connected to Odrenok network.

Now the extensive custom development became an impediment to progress. The alphabetic user interface is primitive by today's standards. The introduction of the PCs as control room machines into the VEPP-4 control system makes programming possible and attractive based on open standards and X display systems.

## 2 BEAM IMAGE OBSERVATION

### 2.1 Layout

Optical diagnostics at the VEPP-3 and the VEPP-4M rings [3] is provided by the dissectors and linear CCD arrays. The optical component of the synchrotron radiation (SR) of the beams is used for this. The dissectors are applied for the measurements of the length and the transversal size of the bunches, and for the study of the collective effects. The linear CCD array provides the measurements of the vertical size of the beams. Dimensions of the beam are: $\sigma_z = 0.1$ mm, $\sigma_r = 1.5$ mm, $\sigma_\varphi = 5$ cm (E = 5 GeV, $U_r$ = 5 MeV). Now the linear CCD arrays are replaced by CCD matrix SONY ICX084AL (659×489 pixels).

The scheme of the optical diagnostic devices is shown in Fig.1.

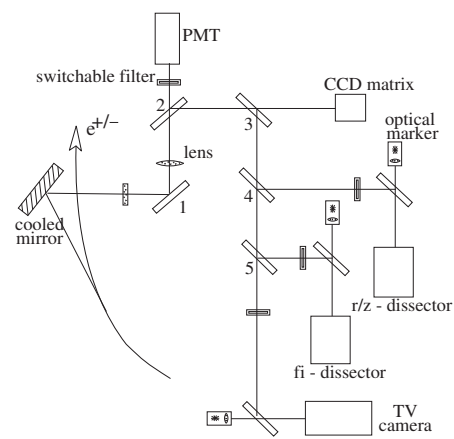

Figure 1: Layout of the optical diagnostic devices.

Two channels of SR output with the optical diagnostics devices are located at the VEPP-4M (each channel is for electron or positron direction only) and one channel is placed at the VEPP-3.

The optical component of SR is reflected by cooled metallic mirror and sent out from the vacuum chamber through a glass window. Mirror 1 matches the SR ray with an optical axis of the system. The beam image is set up on a lens. Mirrors 2-5 split the optical beam to different measurement devices.

## 2.2 Electronics of CCD

The CCD Matrix Controller (MC) (Fig. 2) includes 5 MHz 16-bit ADC and fast Ethernet transmitter. The buffer registers and the interaction logic are based on PLD Altera.

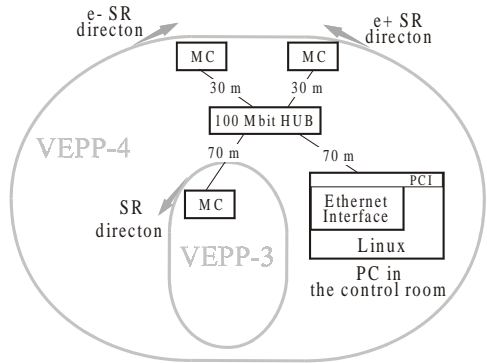

Figure 2: Layout of the beam size observation system at the VEPP4 facility.

The use of the Linux allows to easily connect MCs to the PC via fast Ethernet interface card. The total circumference of the VEPP-4M facility is 366 m therefore intermediate HUB for MCs is used for the connection to the PC.

## 2.3 Software

The use of the CCD matrix instead of the linear CCD array allows to observe a beam spot and to obtain both transverse dimensions $\sigma_z$, $\sigma_r$ and transverse beam particles density. The high speed data transmission provides up to 10 images per second on the computer display.

A custom protocol for the transfer of control to MC and data acquisition had been developed. This protocol provides an on-line exchange between MC and the PC. Each exchange operation includes sending of the control packet from the PC with function words for controller and status words for ADC, and receiving of the data packets sequence from MC. Each data packet from MC has length of 1.3 kByte including one string of the matrix, therefore the total number of the packets of the beam image is about 500. It takes about 100 ms to receive and process the beam image in the PC. Image processing in the PC is performed with ROOT framework.

Now this system is in preliminary usage. First test pictures are available at this site[1].

## 3 BEAM ENERGY MEASUREMENT

### 3.1 Layout

For carrying out the experiment of the precision measurement of the τ-lepton mass near its production threshold at 1.8 GeV, it is necessary to know exactly the energy of colliding particles. For this purpose, special systems including counters of scattered electrons and plates of the TEM wave depolarizer are installed on the booster storage ring VEPP-3 and on the collider VEPP-4M [4].

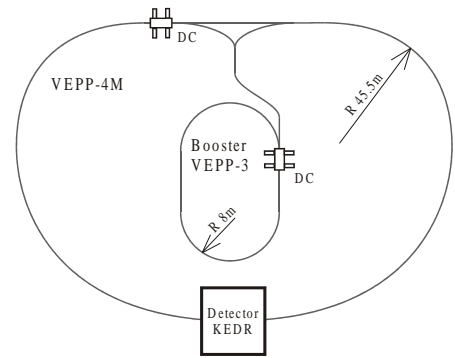

Figure 3: Layout of the Depolarizer-Counters.

The layout of the Depolarizer-Counter (DC) devices is shown in Fig.3. The current hardware configuration of the measuring system is depicted in Fig. 4.

The system consists of one PC, three CAMAC crates, connected to the PC via special serial link, and 8 stepper-motor controllers, connected to the PC via RS-485 serial link.

The PC is physically located in the main control room. It provides the operator interface to the control and measuring system.

All CAMAC crates are located in the radio control room. Two CAMAC crates include counter electronics, where all signals from counters are concentrated. One CAMAC crate includes the frequency synthesizer with the minimal band width of $\Delta f_d \sim 1 \div 10$ Hz and the rearrangement step of 1 Hz.

The Stepper-Motor Controllers (SMC) are remotely located at the facility's two DCs (see Fig. 3 ).

---

[1] http://www.vepp4-pult1.inp.nsk.su/~vepp4/size4/index.html

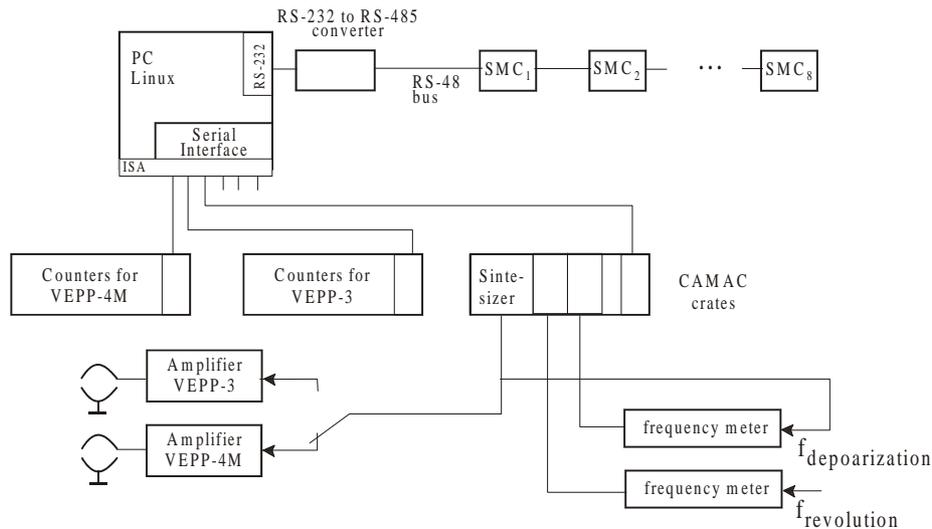

Figure 4: The layout of the hardware.

## 3.2 Software

The software for the measurement system is based on the ROOT package. The ROOT system was developed in CERN, and provides a set of frameworks sufficient to handle and analyze data.

This software controls the frequency synthesizer and the stepper motors with X-window based tools, which prompts the operator to fill in the necessary values. Fields in the windows can be modified and used values are displayed.

The operator can observe a visual representation of the data acquisition from the counters in different graphical windows. He can interactively process the data with curve fitting, minimization, graphics and visualization tools.

The software is linked with the VEPP-4 control system [2] for getting the status of the VEPP-4 facility, the beam energy and the beam current values.

Some pictures of the depolarization process observation are available at this site[2].

## 4 CONCLUSION

There is no possibility to abandon CAMAC and Odrenok computers in a short time. One of the solutions is to expand of the usage GUI based applications for new control tasks. This approach will enhance the flexibility of the accelerator operations and provide new options for the accelerator control. The main goal of the future development of the VEPP-4 control system is to entirely retire the Odrenok computers.

---

[2] http://www.vepp4-pult1.inp.nsk.su/~vepp4/polar/index.html